\documentclass{elsart}
\usepackage{epsfig}
\usepackage{amssymb}
\usepackage{lscape}
\usepackage{amsmath}
\usepackage{bm}
\usepackage{graphicx}
\usepackage{graphics}
\usepackage{color}
\usepackage{pstricks}
\usepackage{cite}
\usepackage[leftbars]{changebar}
\newcommand{\oI}{\Omega \times I}
\newcommand{\bDelta}{{\bm \Delta}}
\newcommand{\Rey}{\textrm{Re}}
\newcommand{\bnabla}{{\bm \nabla}}
\newcommand{\ddt}[1]{\frac{\partial #1}{\partial t}}

\newcommand{\uu}{{\mathbf u}}

\newcommand{\xx}{{\mathbf x}}

\makeatletter
\DeclareRobustCommand{\Cpp}
{\valign{\vfil\hbox{##}\vfil\cr
   \textsf{C\kern-.1em}\cr
   $\hbox{\fontsize{\sf@size}{0}\textbf{+\kern-0.05em+}}$\cr}%
}
\makeatother

\makeatletter
\DeclareRobustCommand{\Cp}
{\valign{\vfil\hbox{##}\vfil\cr
   \textsf{C\kern-.1em}\cr
   $\hbox{\fontsize{\sf@size}{0}\textbf{\kern-0.05em}}$\cr}%
}
\makeatother
\journal{Parallel Computing}
\begin{document}

\begin{frontmatter}
\title{Computational performance of a parallelized\\high-order spectral and mortar element toolbox}
\author[EPFL]{Roland Bouffanais\corauthref{cor}\thanksref{FNRS}},
\corauth[cor]{Corresponding author.}
\thanks[FNRS]{Supported by a Swiss National Science Foundation Grant No. 200020--101707}
\ead{roland.bouffanais@epfl.ch}
\author[EPFL]{Vincent Keller},
\ead{vincent.keller@epfl.ch}
\author[EPFL]{Ralf Gruber},
\ead{ralf.gruber@epfl.ch}
\author[EPFL]{Michel O. Deville}
\ead{michel.deville@epfl.ch}
\address[EPFL]{Laboratory of Computational Engineering,\\ \'Ecole Polytechnique F\'ed\'erale de Lausanne,\\ STI -- ISE -- LIN, Station 9,\\ CH--1015 Lausanne, Switzerland}
\begin{abstract}
In this paper, a comprehensive performance review of a MPI-based high-order spectral and mortar element method \Cpp{} toolbox is presented. The focus is put on the performance evaluation of several aspects with a particular emphasis on the parallel efficiency. The performance evaluation is analyzed and compared to predictions given by a heuristic model, the so-called $\Gamma$ model. A tailor-made CFD computation benchmark case is introduced and used to carry out this review, stressing the particular interest for commodity clusters. Conclusions are drawn from this extensive series of analyses and modeling leading to specific recommendations concerning such toolbox development and parallel implementation.
\end{abstract}
\end{frontmatter}

% ===================================================
\section{Introduction}

This paper provides a detailed performance evaluation of the \Cpp{} toolbox named Speculoos (for \underline{Spec}tral \underline{U}nstructured E\underline{l}ements \underline{O}bject-\underline{O}riented \underline{S}ystem). Speculoos is a spectral and mortar element analysis toolbox for the numerical solution of partial differential equations and more particularly for solving incompressible unsteady fluid flow problems \cite{kemenade96:_incoom}. The main architecture choices and the parallel implementation were elaborated and implemented by Van Kemenade and Dubois-P\`elerin \cite{dubois-pelerin99,dubois-pelerin98:_specul}. Subsequently, Speculoos' \Cpp{} code has been growing up with additional layers enabling to tackle and simulate more specific and arduous CFD problems: viscoelastic flows by Fi\'etier and Deville \cite{fietier03:_detec,fietier03:_linear,fietier03:_time}, fluid-structure interaction problems by Bodard and Deville \cite{bodard04:_fluid}, large-eddy simulations of confined turbulent flows by Bouffanais \etal\ \cite{bouffanais05:_large,bouffanais06:_large} and free-surface flows by Bouffanais and Deville \cite{bouffanais05:_mesh_updat_techn_free_surfac}. 

It is well known that spectral element methods are amenable easily to parallelization as they are intrinsically a natural way of decomposing a geometrical domain \cite{fischer91:_paral_stokes} and Chap. 8 of \cite{deville02:_high}.

The numerous references previously given and the ongoing simulations based on Speculoos highlight the achieved versatility and flexibility of this \Cpp{} toolbox. Nevertheless, ten years have passed between the first version of Speculoos' code and now, and tremendous changes have occurred at both hardware and software levels: fast dual DDR memory, RISC architectures, 64-bit memory addressing, compilers improvement, libraries optimization, libraries parallelization, increase in inter-connecting switch performance, etc. 

Back in 1995, Speculoos was commonly compiled and was running on HP, Silicon Graphics workstations and also on the Swiss-Tx machine, a commodity-technology based computer with enhanced interconnect link between processors \cite{gruber97:_swiss_tx}. Currently most of the simulations based on Speculoos are compiled and are running on commodity clusters. The workstation world experienced a technical revolution with the advent of `cheap' RISC processors leading to the ongoing impressive development of parallel architectures such as massively parallel clusters and commodity clusters. As a matter of fact, Speculoos benefited from this fast technical evolution as it was originally developed as to run in a single program, multiple data mode (SPMD) on a distributed-memory computer. The performance evaluations presented here are demonstrating the correlation between the good performances measured with Speculoos and the adequation of this code structure with the current hardware and software evolutions in parallel computing.

This paper is organized as follows. In Section \ref{sec:spec-numer-cont} we introduce the numerical context in which Speculoos was initiated, the software aspects related to its implementation and the variable-size benchmark test case used for the performance evaluation presented in the subsequent sections. Section \ref{sec:parallel-implementation} is devoted to the parallel performance analysis achieved on RISC-based commodity clusters. Finally, in Section \ref{sec:conclusions} we draw some conclusions on the results obtained.

% ===================================================
\section{Speculoos numerical and software context}\label{sec:spec-numer-cont}

In this section, is gathered the necessary background information regarding the numerical method---namely the spectral and mortar element method---, the object-oriented concept and the parallel paradigm, essential roots embodied in Speculoos. The final Section \ref{sec:benchm-eval-test} introduces the simulation used throughout this study as benchmark evaluation test case.

%-----------------------------------------------
\subsection{Spectral and mortar element method}\label{sec:spectr-mort-elem}

The spectral element method (SEM) is a high-order spatial discretization method for the approximate Galerkin solution of partial differential equations expressed in weak forms. The SEM relies on expansions on Lagrangian interpolants bases used in conjunction with particular Gauss--Lobatto and Gauss--Lobatto--Jacobi quadrature rules \cite{maday89:_spect_navier_stokes,patera84:_spect}. As high-order finite element techniques, the SEM can deal with arbitrary complex geometry where $h$-refinement is achieved by increasing the number of spectral elements and $p$-refinement by increasing the Lagrangian polynomial order within the elements. From a high-order precision viewpoint, SEM is comparable to spectral methods as an exponential rate-of-convergence is observed when smooth solutions to regular problems are sought.

${\mathcal C}^0$-continuity across element interfaces requires the exact same interpolation in each and every spectral element sharing a common interface. The associated caveat to such conforming configurations is the over-refinement meshing generated in low-gradient zones. The adopted remedy to such nuisance is a technique developed by Bernardi \etal\ \cite{bernardi94} referred to as the mortar element method. Mortars can be viewed as variational patches of the discontinuous field along the element interfaces. They relax the ${\mathcal C}^0$-continuity condition while preserving exponential rate-of-convergence, and thus allow polynomial nonconformities along element interfaces.

%-----------------------------------------------
\subsection{Parallel implementation}\label{sec:parallelism}

The complexity and the size of the large three-dimensional problems tackled by numericists in their simulations require top computational performance accessible from highly parallelized algorithms running on parallel architectures. As mentioned in \cite{dubois-pelerin99}, the implementation of concurrency in Speculoos was based on the concept that concurrency is a painful implementation constraint going against the high-level object-oriented programming concepts. As a matter of consequence, Speculoos parallelization was kept very low-level. In most higher-level operations parallelism does not even show up. 

From a computational viewpoint, systems discretized with a high-order spectral element method rely mainly on optimized tensor-product operations taking place at the spectral element level. The natural data distribution for high-order spectral element methods is based on an elemental decomposition in which the spectral elements are distributed to the processors available for the run. It is worth noting that for very large computations, the number of spectral elements can become relatively important as compared to the number of processors available for the computation. The design of Speculoos makes it possible to have several elements sitting on a single processor. Nodal values on subdomain interface boundaries are stored redundantly on each processor corresponding to the spectral elements having this interface in common. Moreover, this approach is consistent with the element-based storage scheme which minimizes the inter-processor communications. Inter-processor communication is completed by MPI instructions \cite{gropp99:_using_mpi}.

%-----------------------------------------------
\subsection{Benchmark evaluation test case description}\label{sec:benchm-eval-test}

As a common practice in performance evaluation, it is important to build a tailor-made benchmark based on a numerical simulation corresponding to a concrete situation. Before proceeding to the first step of our performance evaluation, we have short-listed some key parameters that have the most significant impact on the performance of our toolbox: single-processor optimization on the three computer architectures described in Table \ref{tab:machines}, single-processor profiling analysis, parallel implementation and scalability (including speedup, efficiency, communication times) and parallel implementation and processor dispatching. 

A test case has been developed for this benchmark and for the parallel benchmarking, see Sec. \ref{sec:parallel-implementation}. This test case belongs to the field of CFD and consists in solving the Navier--Stokes equations for a viscous Newtonian incompressible fluid. Based on the problem at hand, it is always physically rewarding to non-dimensionalize the governing Navier--Stokes equations which take the following general form
\begin{align}
\ddt{\uu} + \uu \cdot \bnabla \uu &= - \bnabla p+ \frac{1}{\Rey} \bDelta \uu + \textrm{\bf f} , & & \forall (\xx, t) \in \oI\label{eq:chap2:NS},\\ 
\bnabla \cdot \uu &= 0, & & \forall (\xx, t) \in \oI \label{eq:chap2:divv},
\end{align}
where $\uu$ is the velocity field, $p$ the reduced pressure (normalized by the constant fluid density), $\textrm{\bf f}$ the body force per unit mass and $\textrm{Re}$ the Reynolds number expressed as
\begin{equation}\label{eq:chap2:Re}
\textrm{Re}=\frac{UL}{\nu},
\end{equation}
in terms of the characteristic length $L$, the characteristic velocity $U$ and the constant kinematic viscosity $\nu$. The system evolution is studied in the time interval $I=[t_0,T]$. From the physical viewpoint, Eqs. \eqref{eq:chap2:NS}--\eqref{eq:chap2:divv} are derived from the conservation of momentum and the conservation of mass respectively. For incompressible viscous fluids, the conservation of mass also called continuity equation, enforces a divergence-free velocity field as expressed by Eq. \eqref{eq:chap2:divv}. Considering particular flows, the governing Navier--Stokes equations \eqref{eq:chap2:NS}--\eqref{eq:chap2:divv} are supplemented with appropriate boundary conditions for the fluid velocity $\uu$ and/or for the local stress at the boundary. For time-dependent problems, a given divergence-free velocity field is required as initial condition in the internal fluid domain.

All our computations were carried out using two time integrators: the implicit backward-differentiation formula (BDF) of order 2 for the treatment of the Stokes operator and an extrapolation scheme (EX) \cite{couzy95:_spect_elemen_discr_unstead_navier,karniadakis91:_high_navier} of same order for the nonlinear convective term. One type of pressure decomposition mode, based on a fractional-step method using pressure correction namely {\tt BP1-PC} \cite{couzy94:_spect_uzawa,perot93,perot95:_commen} is used.

Speculoos uses a Legendre SEM \cite{maday89:_spect_navier_stokes,patera84:_spect,deville02:_high} for the spatial discretization of the Navier--Stokes equations. For the sake of simplicity the same polynomial order has been chosen in the different spatial directions ($N_x=N_y=N_z=N$). Moreover, to prevent any spurious oscillations in our Navier--Stokes computations, the choice of a staggered ${\mathbb P}_N - {\mathbb P}_{N-2}$ interpolation method for the velocity and pressure respectively, has been made \cite{deville02:_high,maday92:_nimes_n_stokes}. As a consequence of this choice of a staggered grid, the inner-element grid for the $x$-, $y$- and $z$-component of the velocity field is a Gauss--Lobatto--Legendre grid made up with $(N+1)^2$ quadrature (nodal) points and the grid for the pressure is a Gauss--Legendre grid made up with $[(N-2)+1]^2$ quadrature (nodal) points, in each spectral element.

The test case corresponds to the fully three-dimensional simulation of the flow enclosed in a lid-driven cubical cavity at the Reynolds number of $12\,000$ placing us in the locally-turbulent regime. It corresponds to the case denoted under-resolved DNS (UDNS) in Bouffanais \etal\ \cite{bouffanais05:_large,bouffanais06:_large}. The reader is referred to Bouffanais \etal\ \cite{bouffanais05:_large,bouffanais06:_large} for full details on the numerical method and on the parameters used throughout the present paper.

% ===================================================
\section{Parallel implementation}\label{sec:parallel-implementation}

In the sequel, we will assume that the reader is familiar with the basics of parameterization on a parallel machine. For a complete introduction to these notions we refer the reader to the following references \cite{gruber03:_param,gruber04:_scalab_aspec_commod_clust}.

The speedup $S$ of an application on a given parallel machine can be described as
\begin{equation}
S= \frac{\mbox{Computing time on one processor}}{\mbox{CPU plus communication times on }P\mbox{ processors}} = \frac{T_1}{T_P+T_C}.
\end{equation}
If we suppose that the computing effort strictly scales with $P$, then $T_1=PT_P$ and the speedup can be written as
\begin{equation}\label{eq:app3:Gamma}
S=\frac{T_1}{T_P+T_C}=\frac{PT_P}{T_P+T_C}=\frac{P}{1+\gamma_m/\gamma_a}=\frac{P}{1+1/\Gamma},
\end{equation}
where
\begin{equation}
\gamma_a = \frac{\text{number of operations [MFlop]}}{\text{amount of data to transfer [MWord]}},
\end{equation}
is related to the application and
\begin{equation}
\gamma_m = \frac{\text{effective processor performance [MFlops]}}{\text{effective communication bandwidth per processor [MWords]}},
\end{equation}
to the machine, and $\Gamma = \gamma_a /\gamma_m$. The reader is referred to \cite{gruber03:_param} for full details on such parameterisation to tailor commodity clusters to applications. The efficiency $E$ of a parallel machine is defined by 
\begin{equation}
E=\frac{S}{P}=\frac{1}{1+1/\Gamma}.
\end{equation}

%-----------------------------------------------
\subsection{Speculoos characteristics}

Speculoos uses a small amount of main memory. Parallelization is made in order to reduce the high overall computing time. The number of elements and the polynomial degrees in the three space directions are denoted by $E_x$, $E_y$, and $E_z$, and $N_x$, $N_y$, and $N_z$, respectively. The total number of independent variables per element is therefore $n_v\times (N_x+1)\times (N_y+1)\times (N_z+1)$, where $n_v$ is the number of vector components per Gauss--Lobatto--Legendre (GLL) quadrature point. In addition, there are $E_x\times E_y\times E_z$ elements.

%-----------------------------------------------
\subsection{Hardware and software used}

To perform the Speculoos code benchmark, the machines presented in Table~\ref{tab:machines} have been used.

\begin{table}[htbp]
\begin{center}
\begin{tabular}{lccccc}
  \emph{\bf Name} & \emph{\bf Manufacturer} & \emph{\bf CPU type} & \emph{\bf 
  Nodes} & \emph{\bf Cores} & \emph{\bf Interconnect}   \\
\hline \hline
\texttt{Gele}       & Cray   & Opteron DC & 16  & 32  & SeaStar \\
\texttt{Pleiades}   & Logics & Pentium 4  & 132 & 132 & FE \\
\texttt{Pleiades2}  & Dell   & Xeon       & 120 & 120 & GbE \\
\texttt{Pleiades2+} & Dell   & Xeon 5150  & 99  & 396 & GbE \\
\hline
\end{tabular}
\caption{\label{tab:machines}Characteristics of the machines used for the benchmark. DC=Dual-Core. FE=Fast Ethernet. GbE=Gigabit Ethernet.}
  \end{center}
\end{table}

As mentioned previously, the Speculoos code is written in \Cpp{}, uses BLAS operations and implements the Message Passing Interface (MPI). 

The PAPI (Performance API) \cite{papi} available on the Cray XT3 machine was used to measure the number $O$ of operations (in GFlops) and the MFlops rate of Speculoos. The VAMOS service available on the three Pleiades clusters \cite{vamos} maps the hardware related data from the Ganglia monitoring tool \cite{ganglia} with the application and user related data (from cluster Resource Management System and Scheduler). We used the most aggressive optimization flag on all machines (\texttt{-O3} flag).

%-----------------------------------------------
\subsection{Fixed problem size}

\begin{figure}[htbp]
        \begin{center}
                \includegraphics[width=0.7\textwidth]{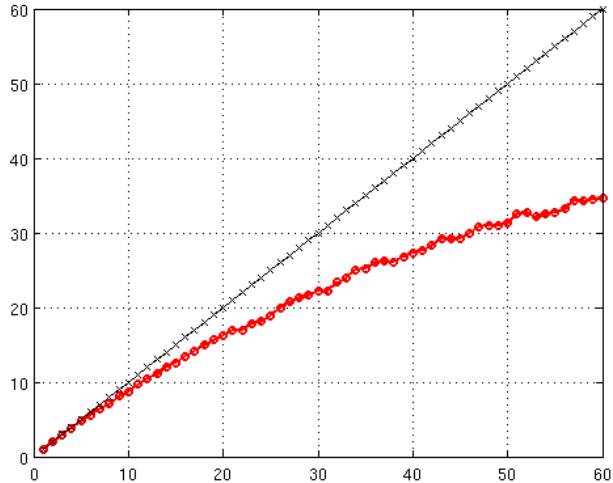} \\
               \caption {\label{fig:speedupP2} Speedup of Speculoos code on the \texttt{Pleiades2} (Xeon CPU).}
        \end{center}
\end{figure}

The first measurements are done on \texttt{Pleiades2} with a fixed problem size, $E_x = E_y = E_z = 8$; $N_x = N_y = N_z = 8$; $O = 155.4$ GFlops, and varying the number $P$ of processing elements from 1 to 32. The evolution of the runtime (for one time-step), the associated MFlops rate, and the efficiency $E$ are given in Table~\ref{tab:FixedProblemSize}. The speedup $S$ as a function of the number of processors is plotted in Fig.~\ref{fig:speedupP2}. One observes that with 8 processors a speedup of 7 can be reached and a speedup of 30 with 45 processors. 

\begin{table}[htbp]
  \begin{center}
    \begin{tabular}{lccccc}
      \emph{P} & \emph{\bf GFlops} & \emph{\bf Runtime (1 step)} & \emph{E} \\
      \hline \hline
      1&       0.638  &       243.59 & 1.00 \\
      2&       1.251  &       124.23 & 0.98 \\
      3&       1.901  &       81.75  & 0.99 \\
      4&       2.395  &       64.88  & 0.94 \\
      5&       3.038  &       51.15  & 0.95 \\
      6&       3.566  &       43.58  & 0.93 \\
      7&       4.101  &       37.89  & 0.92 \\
      8&       5.590  &       34.52  & 0.88 \\
      16&      8.346  &       18.62  & 0.82 \\
      32&     14.179  &       10.96  & 0.70 \\
      \hline
    \end{tabular}
    \caption{\label{tab:FixedProblemSize}Evolution of GFlops rate and runtime for fourth time-step.  $E$: Efficiency.}
  \end{center}
\end{table}

%-----------------------------------------------
\subsection{Increase CPU performance}

In this section, the number of processors on a Cray XT3 is kept fixed at the value $P=4$. Then, we modify the polynomial degree and measure the MFlops rate. The MFlops rate performance metric for each process element is shown on Table~\ref{tab:ModifiedProblemSize}. It increases as the problem size increases. As expected, one deduces that there is a limit on the number of processors that should be used in parallel.  

\begin{table}[htbp]
  \begin{center}
    \begin{tabular}{ccccccc}
      \emph{\bf $E_x - E_y - E_z$} & \emph{\bf $N_x - N_y - N_z$} & 
      \emph{\bf MFlops}  & \emph{\bf Walltime} \\
      \hline \hline
      $8 - 8 - 8$ &   $6 - 6 - 6$    &        1624   &        18.54 \\
      $8 - 8 - 8$ &   $7 - 7 - 7$    &        2580   &        29.79 \\
      $8 - 8 - 8$ &   $8 - 8 - 8$    &        3100   &        50.07 \\
      $8 - 8 - 8$ &   $9 - 9 - 9$    &        3700   &        83.12 \\
      $8 - 8 - 8$ &   $10 - 10 - 10$ &        4150   &        146.97 \\
      $8 - 8 - 8$ &   $11 - 11 - 11$ &        4390   &        257.36 \\
      \hline
    \end{tabular}
    \caption{\label{tab:ModifiedProblemSize}Evolution of MFlops rate and runtime for one time-step on 4 Cray XT3 dual-CPU nodes as a function of the polynomial degree.}
  \end{center}
\end{table}

%-----------------------------------------------
\subsection{Varying the number of processing element $P$ with problem size}

A more common way to measure scalability, and to overcome Amdahl's law, is to fix the problem size per processor and to increase the number of processors with the overall problem size. In other words, one tries to fix $\Gamma$ that measures the ratio between processor needs over communication needs. We show in Table~\ref{tab:samesize} the scalability of Speculoos on the \texttt{Pleiades2+} cluster. It was compiled using MPICH2 and \texttt{icc} \Cpp{} compiler version 9.1e. 

\begin{table}[htbp]
  \begin{center}
    \begin{tabular}{ccccccccccc}
      \emph{\bf $E_x - E_y - E_z$} & \emph{\bf $N_x - N_y - N_z$}    
      & \emph{\bf Nodes-Cores} & \emph{\bf Elem/Core} & \emph{\bf Walltime} \\
      \hline \hline
      $4 - 4 - 4$ &   $8 - 8 - 8$ &   1 - 1 & 64 &    8.68 \\
      $8 - 8 - 8$ &   $8 - 8 - 8$ &   2 - 8 & 64 &    39.26 \\
      $16 - 16 - 16$ &$8 - 8 - 8$ &   16 - 64 &64 &   147.97 \\
      \hline
    \end{tabular}
    \vspace{0.20cm}
    \\
    {\bf (A)}
    \\
    \vspace{0.20cm}
    \begin{tabular}{ccccccccccc}
      \emph{\bf $E_x - E_y - E_z$} & \emph{\bf $N_x - N_y - N_z$}    
        & \emph{\bf Nodes-Cores} & \emph{\bf Elem/Core} & \emph{\bf Walltime} \\
        \hline \hline
        $4 - 4 - 4$ &   $8 - 8 - 8$ &   1 - 1 & 64 &    8.68 \\
        $8 - 8 - 8$ &   $8 - 8 - 8$ &   4 - 8 & 64 &    33.50 \\
        $16 - 16 - 16$ &$8 - 8 - 8$ &   32 - 64 &       64 &    111.71 \\
        \hline
      \end{tabular}
      \vspace{0.20cm}
      \\
      {\bf (B)}
      \\
      \vspace{0.20cm}
      \caption{\label{tab:samesize}Scalability of Speculoos. Same polynomial degree, same number of elements on each computing node on \texttt{Pleiades2+} (Woodcrest) cluster. {\bf (A):} with 4 MPI threads per node. {\bf (B):} with $npernode = 2$ , two MPI threads per node.}
    \end{center}
\end{table}

Table~\ref{tab:samesize} (A) shows results obtained when all the 4 cores are active for $P>1$. Note that one Woodcrest node with 2 dual-core processors (Table~\ref{tab:samesize}) is slightly faster than 4 dual-CPU nodes (Table~\ref{tab:ModifiedProblemSize}) of the Cray XT3. When increasing the number of nodes with the problem size, the MFlops rate per core remains the same for all the cases. At this point, it is legitimate to determine if Speculoos is memory or processor bound. To find out, all the test cases in Table~\ref{tab:samesize} have been resubmitted to the Woodcrest nodes, first (A) using all the 4 cores per node, then (B) restricting to two the maximal number of MPI threads per node. Thus, instead of 16 nodes, 32 nodes were used to run the 64-processor case (see Table~\ref{tab:samesize} (B)). One sees that the overall CPU time has been reduced by 20\%, but the number of nodes was doubled. This shows that Speculoos includes parts that are processor bound and others that are memory bound. As a consequence, using all 4 cores does not give a two fold speedup (as one expects for a processor bound program) but neither the speedup is zero (as for a main memory bound application). Therefore, it is always more efficient to run Speculoos on all the 4 cores per node. 

%-----------------------------------------------
\subsection{CPU usage and the $\Gamma$ model}

CPU usage has been monitored by the VAMOS monitoring service~\cite{vamos} available on the Pleiades clusters. It provides information on the application's behavior. The higher the CPU usage is, the better the machine fits to the application. To perform that monitoring we took the same problem size ($E_x = E_y = E_z = 8$ and $N_x = N_y = N_z = 8$) during the same computing duration (10 hours = 36'000 seconds). The application is run for 10 hours and the number of time-steps performed during this time is counted. With such a methodology, we ensure that each sample can perform a maximum of calculations in a given amount of time. It is equivalent to set the same number of iteration for each sample and to measure the walltime.

\begin{figure}[htbp]
  \begin{center}
    \includegraphics[width=0.6\textwidth]{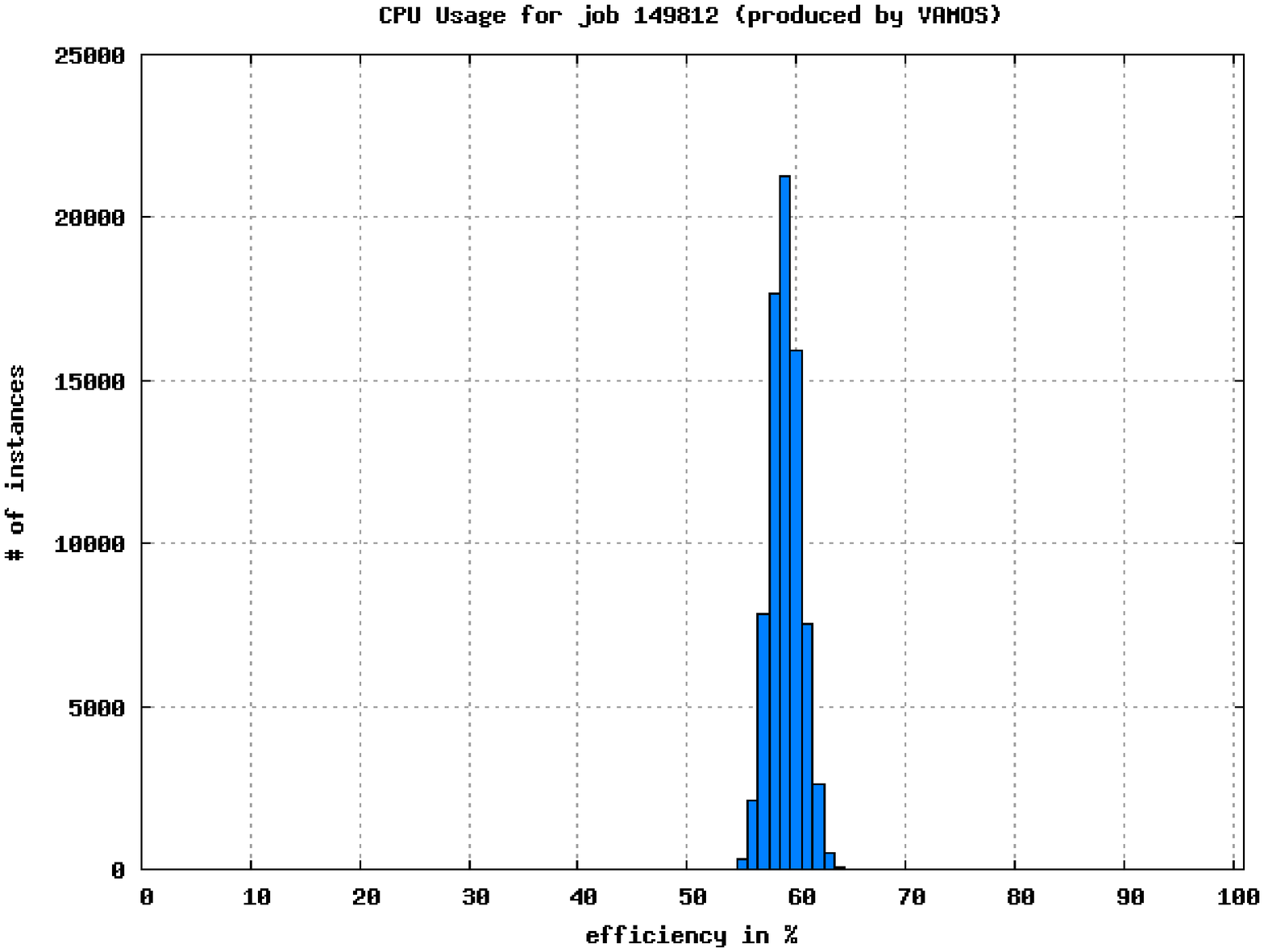} \\
    \includegraphics[width=0.6\textwidth]{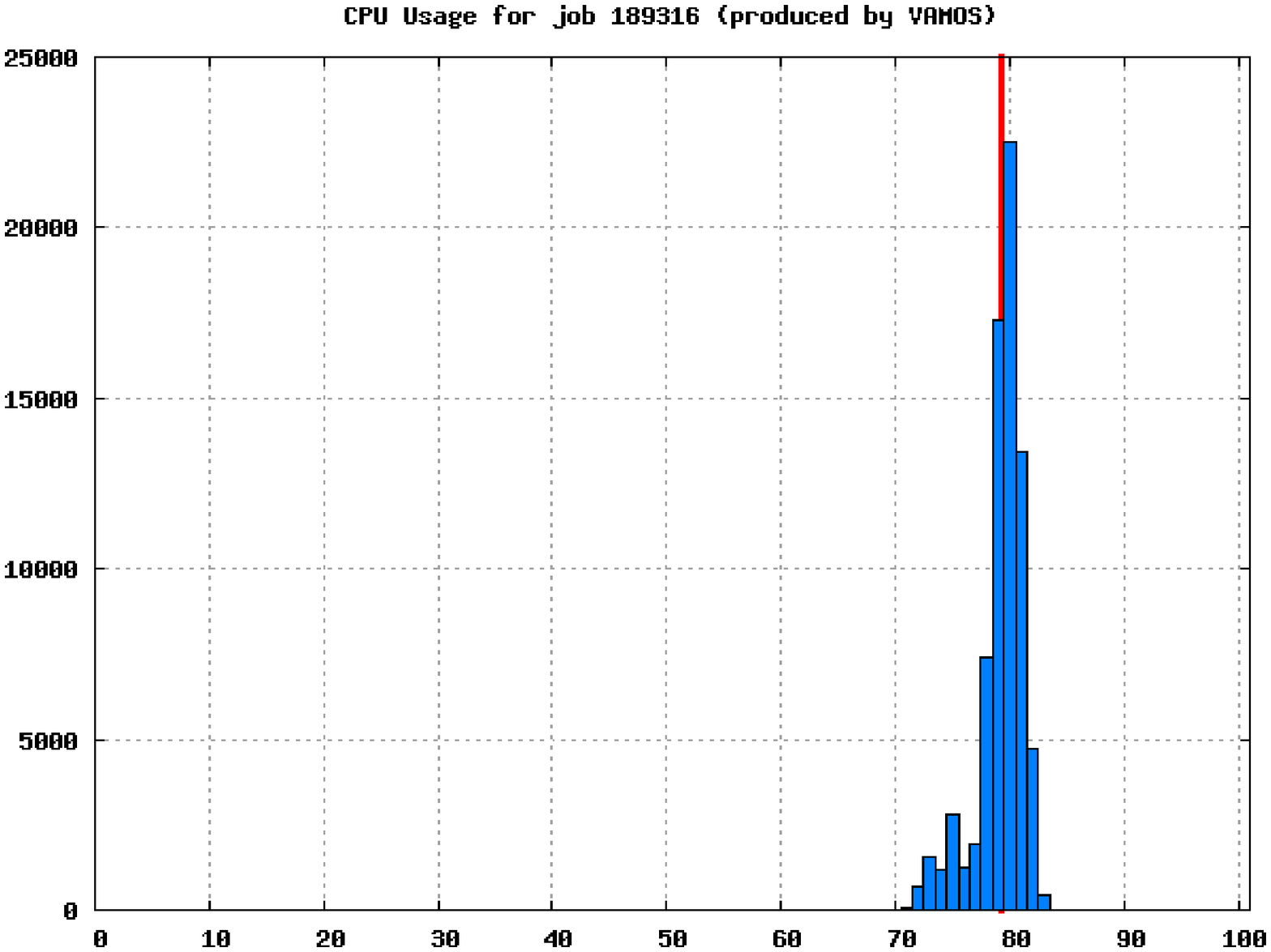} \\
    \includegraphics[width=0.6\textwidth]{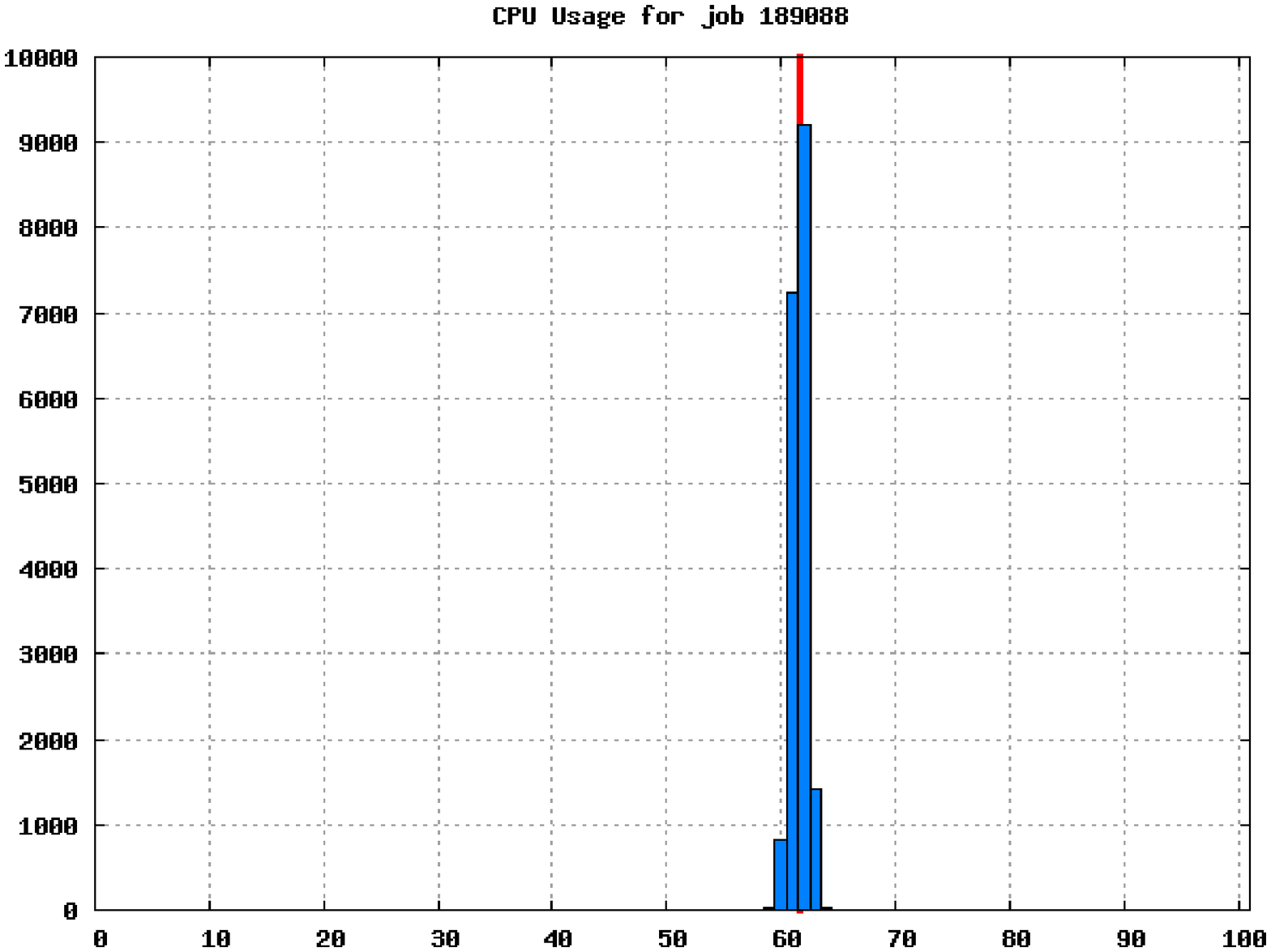} \\
    \caption {\label{fig:p1} CPU Usage of Speculoos on different machines. Top: \texttt{Pleiades} cluster (CPU usage average 51.05\% ,$\Gamma = 1.04$).Middle: \texttt{Pleiades2} cluster 
      (CPU usage average = 79.24\%, $\Gamma = 3.81$). Bottom: \texttt{Pleiades2+} cluster (CPU usage average 61.6\%, $\Gamma = 1.60$).}
  \end{center}
\end{figure}

Figure~\ref{fig:p1} shows the different behavior of Speculoos on the three different \texttt{Pleiades} architectures. The $\Gamma$ value---introduced in Eq. \eqref{eq:app3:Gamma} and, which reflects the ``fitness'' of a given application on a given machine ~\cite{gruber03:_param}---is also computed. Results are reported in Table \ref{tab:Gamma}.

Using the notations introduced earlier, $T$, $T_P$, $T_C$, and $T_L$ denote the total walltime, the CPU time for $P$ processing elements, the time to communicate, and the latency time per iteration step, respectively. Then,
\begin{equation}
T = T_P + T_C + T_L,
\end{equation}
and the parameter $\Gamma$ is easily expressed as
\begin{equation}
\Gamma = \frac{T_P}{T_C+T_L}.
\end{equation}

\begin{table}[htbp]
  \begin{center}
    \begin{tabular}{lcccccccc}
      \emph{\bf } & \emph{\bf T [s]} & \emph{\bf $\Gamma$}  & \emph{\bf b [MB/s]} 
      & \emph{\bf W [words]} & \emph{\bf $T_P$ [s]} & \emph{\bf $T_C$ [s]} &
      \emph{\bf $T_L$ [s]}\\
      \hline \hline
      \texttt{Pleiades} & 23.01* & 1.44* & 12* & $180*10^6$ & 13.58 & 8.43& 1\\
      \texttt{Pleiades2} & 9.55* & 3.81* & 101 & $180*10^6$ &7.56& 0.98&1\\
      \texttt{Pleiades2+} & 12.89* & 1.60* & 101 & $180*10^6$ &7.93& 3.96&1\\
      \hline
    \end{tabular}
    \caption{\label{tab:Gamma}Measured (*) and computed quantities using the $\Gamma$ model.}
  \end{center}
\end{table}

It is possible to measure the total time $T$ by means of an interpretation of the CPU usage plots (see Fig.~\ref{fig:p1}). Indeed, the middleware Ganglia determines for every time interval of 20 seconds the average CPU usage (or efficiency $E$) for each processing element. This information has to be put into relation to the Speculoos application. This is done via the middleware VAMOS. In the plots in Fig.~\ref{fig:p1}, are added up all the values of $E$ that lie in between $x$ and $x+0.01$, where $x$ is the percentile represented on the abscissae of the plots. The efficiency $E$ is related to the $\Gamma$ through 
\begin{equation}
\Gamma = \frac{E}{1-E}.
\end{equation}
What can also be estimated are the network bandwidths $b$ of the GbE switch (between $b=90$ and 100 MB/s per link), the network bandwidth of the Fast Ethernet switch (between $b=10$ and 12 MB/s per link) and the latency ($L = 60\ \mu$s for both networks). First, a consistency test of those quantities is performed. Assuming that the Fast Ethernet switch has a fix bandwidth of $b_1= 12$ MB/s, and for the GbE switch $b_2 = \alpha b_1$, with $\alpha$ unknown. Another unknown is the number of words $W$ that is sent per node to the other nodes, and $T_C = W/b$. Based on the previous assumptions, the three $\Gamma$ values for the three machines and the two networks is expressed as
\begin{align}
\Gamma_1 &=\frac{T_{P_1}}{W/b_1+T_L},\\
\Gamma_2 &=\frac{T_{P_2}}{W/b_2+T_L},\\
\Gamma_3 &=\frac{T_{P_3}}{W/b_2+T_L}.
\end{align}
These constitute a set of three equations for three unknown variables, namely $W$, $\alpha$, and $T_L$. Solving for these variables leads to $T_L=1$, $W=180$ MWords, and $\alpha=8.43$. The value of $b_2=101$ MB/s corresponds precisely to the one measured. This means that the model is well applicable.

%-----------------------------------------------
\subsection{Modification of the number of running threads per SMP node}

\begin{figure}[htbp]
  \begin{center}
    \includegraphics[width=0.6\textwidth]{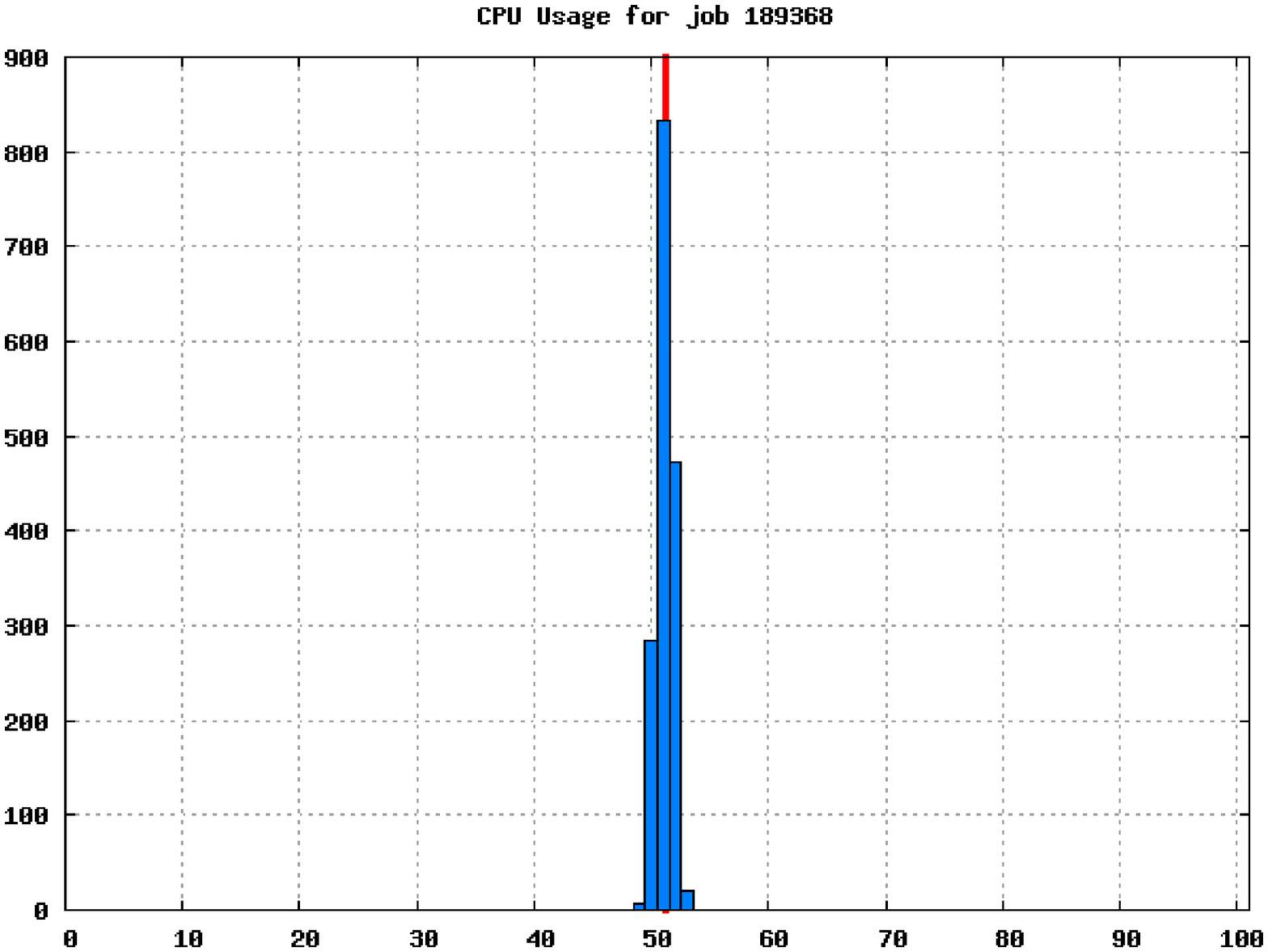} \\
    \includegraphics[width=0.6\textwidth]{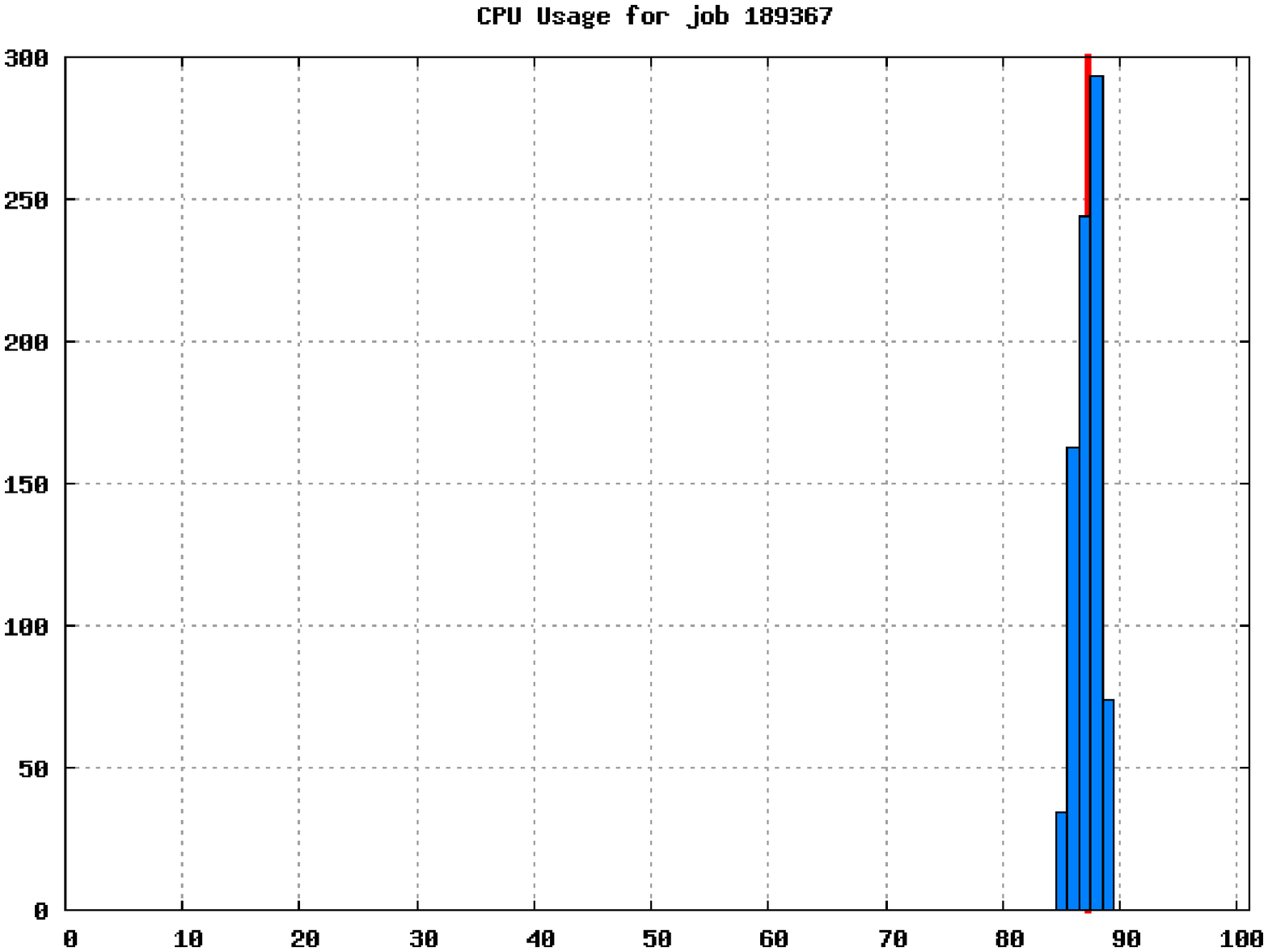} \\
    \caption {\label{fig:p2} CPU Usage on the 5100-series SMP node of \texttt{Pleiades2+} cluster. 16 processing elements were required. 8 nodes/2 cores with 2 MPI threads per nodes in the upper case, 4 nodes/4 cores with 4 MPI threads per node in the lower case.}
  \end{center}
\end{figure}

To study if Speculoos is dominated by inter-node communications, Figure~\ref{fig:p2} shows the result of two runs of the same problem size ($E_x = E_y = E_z = 8$ and $N_x = N_y = N_z = 8$) made respectively on 4 and 8 Woodcrest nodes during the same period of time (1h = 3600 seconds) and counting the number of iteration steps. The first sample was launched forcing 2 MPI threads on each node and the second with 4 MPI threads on each node.

We have to note that the CPU usage (system+user+nice) monitored by Ganglia is the sum of all the process elements. For instance, for a dual-processor machine, when Ganglia measures 50\% CPU usage, it means that each processor run at 100\%. In Figure~\ref{fig:p2}, when 2 MPI threads are blocked per node, we get a CPU usage of 51.13\% while 157 iteration loops have been performed during one hour; when 4 MPI threads run on each node, we get a CPU usage of 87.25\%  while only 117 iteration loops have been performed during one hour. Thus, the real CPU usage for the sample with 2 MPI threads per node is above 100\% (2 cores are unused). 

% ===================================================
\section{Conclusions}\label{sec:conclusions}

The extensive performance review presented in this paper for the high-order spectral and mortar element method \Cpp{} toolbox, Speculoos, has shown that good performances can be achieved even with relatively common internode network communication systems, available software and hardware resources---small commodity clusters with non-proprietary compilers installed on it.

We can conclude that the main implementation choices made a decade ago reveal their promises. Even though those choices could have been questionable ten years ago, they are now in line with the current trend in computer architecture developments with the generalization of commodity and massively parallel clusters. 

The parallel implementation of Speculoos based on MPI has shown to be efficient. Reasonable scalability and efficiency can be achieved on commodity clusters. The results support the original choices made in Speculoos parallel implementation by keeping it at a very low-level. 

One of the goal of this study was to estimate if Speculoos could run on a massively parallel computer architecture comprising thousands of computational units, specifically on the IBM Blue Gene machine at EPFL with 4'096 dual processor units. The performance of one processor corresponds to approximately half of the performance of one processor on the Pleiades commodity cluster. Each Blue Gene node has 512 MB of main memory. A block with $4\times4\times 4$ elements and a polynomial degree of $N=8$ in each space direction takes 200 MB of main memory. In a first step, one block per node will run on one node. Later, Speculoos will be modified to accommodate one block per processor, i.e. two blocks per node. A 4'096 blocks Speculoos case would offer the opportunity to run very accurate simulations of turbulent flows with more than half a billion of unknowns. Such a case would well scale on the IBM Blue Gene solution. In fact, the point-to-point operations per node do not change with the number of nodes. The Gigabit-Ethernet network can well handle the corresponding communications. The all-reduce operations scale logarithmically with the number of computational units. A special efficient Fat Tree network takes care of all multicast communications. As a consequence, large Speculoos cases will perfectly scale on EPFL's Blue Gene machine.

%Acknowledgments
\ack
This research is being partially funded by a Swiss National Science Fundation Grant (No. 200020--101707) and by the Swiss National Supercomputing Center CSCS, whose supports are gratefully acknowledged.

The results were obtained on supercomputing facilities at the Swiss National Supercomputing Center CSCS and on Pleiades clusters at EPFL--ISE.
%\newpage

 \end{document}